\input{aipcheck}

\documentclass[
   ,final            
  ]
  {aipproc}

\layoutstyle{8x11double}

\usepackage{amsmath,amssymb,amstext}
\usepackage{dsfont}
\usepackage{color}

\def\eq#1{(\ref{#1})}

\newcommand {\apgt} {\ {\raise-.5ex\hbox{$\buildrel>\over\sim$}}\ }
\newcommand {\aplt} {\ {\raise-.5ex\hbox{$\buildrel<\over\sim$}}\ }

\def\s0#1#2{\mbox{\small{$ \frac{#1}{#2} $}}}
\def\0#1#2{\frac{#1}{#2}}

\def\dr{{D\!\llap{/}}\,}

\newcommand{\Tr}{\mathrm{Tr}}

\newcommand{\tr}{\mathrm{tr}}

\newcommand{\I}{\mathrm{i}}
\newcommand{\be}{\begin{eqnarray}}
\newcommand{\ee}{\end{eqnarray}}

\newcommand{\Nc}{N_{\text{c}}}

\begin{document}

\title{The QCD phase diagram: Results and challenges}


\author{Jan M.~Pawlowski}{ address={Institut f\"ur Theoretische
    Physik, Universit\"at Heidelberg, Philosophenweg 16, 69120
    Heidelberg, Germany} ,altaddress={ExtreMe Matter Institute EMMI,
    GSI, Planckstr. 1, 64291 Darmstadt,
    Germany} 
}

\begin{abstract}
  I review the progress made in recent years with functional methods
  in our understanding of the QCD phase diagram. In particular I
  discuss a renormalisation group approach to QCD at finite
  temperature and chemical potential. Results
  include the location of the confinement-deconfinement phase
  transition/cross-over and the chiral phase transition/cross-over
  lines, their nature as well as their interrelation.
\end{abstract}

\maketitle


\subsubsection{Introduction} Quantum Chromodynamics (QCD) at finite
temperature and density is a very active area of research. The
equation of state of QCD and the nature of the
transition from the hadronic phase with broken chiral symmetry
to the chirally symmetric, deconfined quark-gluon plasma phase is 
of great importance for a better understanding of the
experimental data, e.g.~\cite{BraunMunzinger:2003zd}.

For full QCD with dynamical quarks one expects the deconfinement phase
transition to be a crossover as quarks explicitly break the underlying
center symmetry of the gauge group. The nature of the chiral phase
transition primarily depends on the value of the current quark mass,
which explicitly breaks chiral symmetry, as well as on the strength of
the chiral anomaly~\cite{Pisarski:1983ms}.  While the confinement
phase transition is driven by gluodynamics, the chiral phase
transition is governed by strong quark interactions. It is a
highly non-trivial observation that both lie remarkably close at least
for small quark chemical
potentials~\cite{Aoki:2006br,deForcrand:2002ci}. An understanding of
this interrelation is subject of an ongoing debate.

Hence, the resolution of the QCD phase diagram requires both, the
computation of the relevant observables, in particular the order
parameters, and also some analytic understanding of the mechanisms
involved, see e.g.~\cite{Alkofer:2010ue}. Such a twofold task is best obtained
within a combination of different methods that allow a direct access
to the observables as much as to the mechanisms behind the observed
phenomena. Functional continuum methods such as functional
renormalisation group equations (FRG) and Dyson-Schwinger equations
(DSE) are well-suited for the above task. They are indeed
complementary to lattice simulations: most importantly they are
amiable towards chiral fermions including the anomalous breaking of
chiral symmetry and are straightforwardly applied to finite chemical
potential. Ideally lattice simulations and functional approaches go
hand in hand, and should be used to improve and to check respective
results, also leading to better systematic error estimates.  In
combination, this should allow us to map out the phase diagram of QCD.

\subsubsection{Functional RG}
We shall present results on the order parameter for the chiral and the
confinement-deconfinement phase transition obtained with functional
methods, mostly with the FRG.  Within a functional approach one
usually computes correlations functions of quarks and gluons or that
of composite operators such as hadronic degrees of freedom. These
correlations functions are related by an infinite hierarchy of partial
integro-differential equations which are solved within specific
approximations to the full system at hand. Clearly a sound discussion
of the approximations is of chief importance for the reliability of
functional approaches. However, it is beyond the scope of the present
overview, and for more details we defer the reader to the original
works.

\begin{figure}[h]
  \includegraphics[height=.065\textheight]{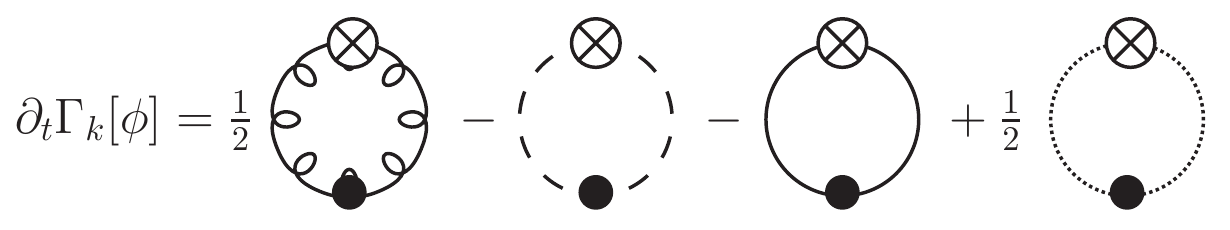}\label{fig:funflowQCD}
  \caption{Functional flow for QCD: the lines denote full field
    dependent propagators. Crosses denote the cut-off insertion
    $\partial_t R$. }
\end{figure}

Most of the results presented here are obtained in Landau gauge
QCD with the classical action
\begin{eqnarray}\nonumber 
  S_{\rm QCD}&=& 
  \014 \int_x {F^a_{\mu\nu}}^2 +  \0{1}{2 \xi} \int_x (\partial A^a)^2 
  +\int_x  \bar C^a (\partial D)^{ab} C^b\\ 
  & &+\int \bar\psi \left( i \dr +i \,m_\psi 
    +i \mu\gamma_0 \right)\psi\,,
\label{eq:SQCD}\end{eqnarray} 
where $\xi\to 0$. An infrared regularisation is achieved with the
introduction of momentum-dependent mass-terms for ghost, gluon and
quark fields. We also introduce cut-off terms for effective low
energy degrees of freedom such as mesons and baryons. This leads to a
scale-dependent effective action, $\Gamma_k[A,C,\bar
C,\psi,\bar\psi;\sigma,\vec \pi,...]$, with infrared cut-off scale
$k$. The $\sigma$ and $\vec \pi$ fields stand for mesonic composite
operators. For $k\to 0$, we approach the fully interacting theory,
whereas for $k\to \infty$ we are left with the asymptotically free
high energy QCD. An infinitesimal change of the scale $k$ is described
by Wetterich's flow equation, \cite{Wetterich:1992yh}, 
\begin{equation}\label{eq:FRG} 
  \partial_t  \Gamma _k [\phi] =\frac{1}{2} \Tr \left(
    \Gamma_k^{(2)}[\phi]+R_k\right)^{-1} \partial_t R_k \,, 
\end{equation}  
where $t=\ln k/\Lambda$, and $\phi=(A,C,\bar
C,\psi,\bar\psi;\sigma,\vec \pi,...)$. Finally, $\Gamma^{(2)}_k$ denotes
the second derivative of $\Gamma_k$ w.r.t. the fields. The cut-off
functions $R_k$ provide infrared cut-offs for all fields, including
the composite fields, i.e. $\sigma$ and $\vec\pi$, for details see the
reviews on gauge theories,
\cite{Litim:1998nf,Pawlowski:2005xe,Gies:2006wv,Igarashi:2009tj}. We
emphasise that \eq{eq:FRG} comprises a first principle QCD flow, the
appearance of composite operators does not signal an effective field
theory setting but rather a convenient parametrisation, see
\cite{Pawlowski:2005xe,Gies:2001nw}. The flow \eq{eq:FRG} has a simple
diagrammatic form depicted in Fig.~\ref{fig:funflowQCD}, see
\cite{Braun:2009gm}.
%
%
The first and second loop generate gluon and ghost fluctuations
respectively, the third loop generates fluctuations of the quarks, and
the last loop stands for the loops of the mesonic $\sigma$- and
$\vec\pi$-fluctuations, and possible further hadronic degrees of
freedom.

\subsubsection{Flows for Yang-Mills propagators in the Landau gauge}
The Landau gauge has very peculiar properties that facilitate the
computation of correlations functions: first of all, the ghost-gluon
vertex is protected from renormalisation. Second, we have infrared
ghost dominance. Its weak form relevant for the present investigations
simply entails that the gluon dressing function is vanishing in the
infrared, $p^2/\Gamma^{(2)}(p\to 0)\to 0$, whereas the ghost dressing
function does not, $p^2/\Gamma_C^{(2)}(p\to 0)>0$. For a detailed
discussion of the global properties of Landau gauge Yang-Mills theory
see \cite{Fischer:2008uz,vonSmekal:2008ws}. A related gauge with
similar properties is Coulomb gauge, for a FRG study and further references see
\cite{Leder:2010ji,Reinhardt:2010xm}.

\begin{figure}[h]
  \includegraphics[height=.28\textheight]{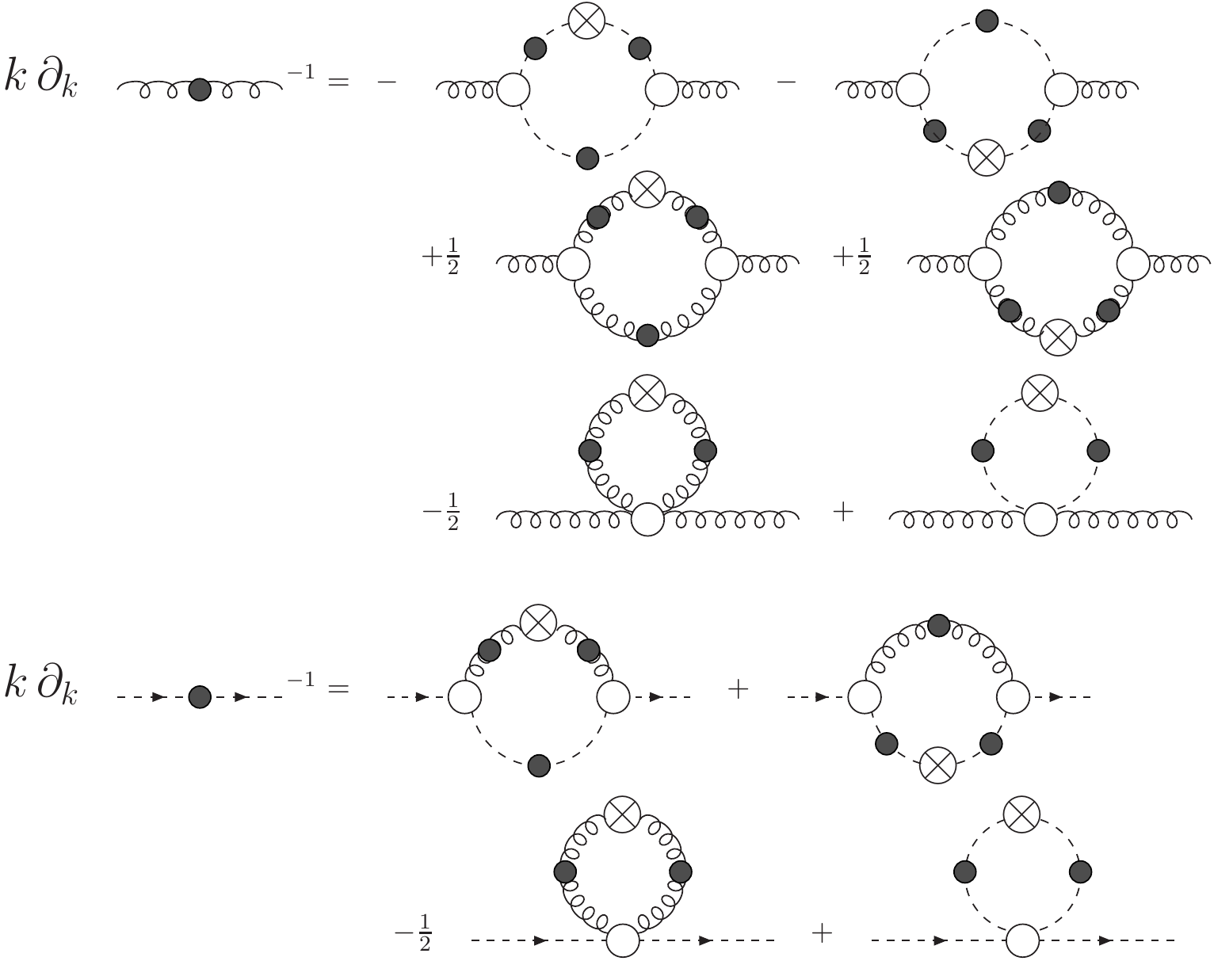}\label{fig:propflow}
  \caption{Functional RG equations for the gluon and ghost propagator.
    Filled circles denote dressed propagators and empty circles denote
    dressed vertex functions. Crosses denote the cut-off insertion
    $\partial_t R$.  \ \vspace{-.4cm}\ }
\end{figure}

\begin{figure}[b]
  \includegraphics[height=.25\textheight]{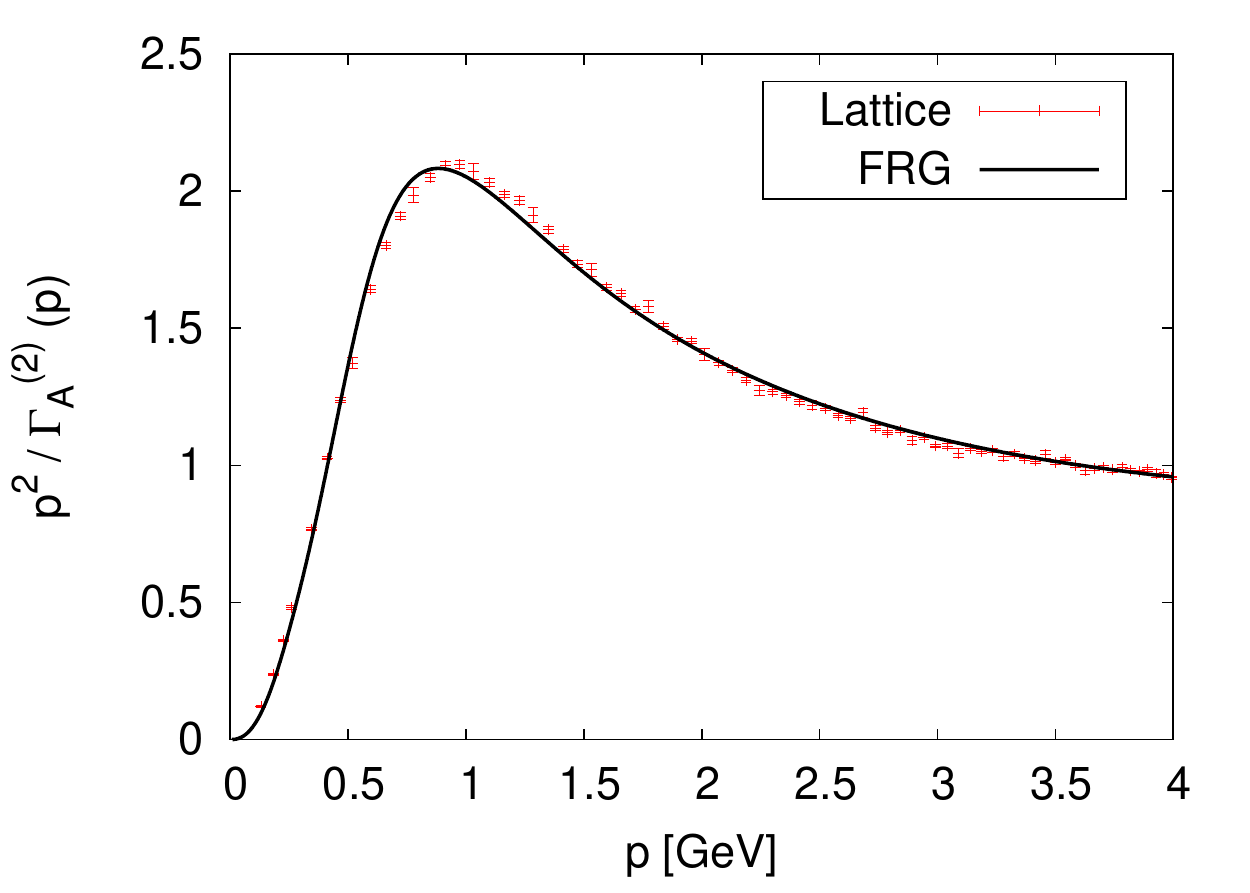}\label{fig:propplots}
  \caption{Gluon dressing function: FRG \cite{Fischer:2008uz},
    lattice \cite{Sternbeck:2006cg}.  \ \vspace{-.4cm}\ }
\end{figure}

The full set of flow equations for ghost and gluon propagators is
derived from Fig.~\ref{fig:funflowQCD} with two derivatives w.r.t. the
ghost fields and the gluons respectively. The propagator flows are
coupled sets of integro-differential one loop equations for the
propagators that also depend on vertex functions with up to four
legs, see Fig.~\ref{fig:propflow}. 
%
%
%
This system of flow equations has been studied in
\cite{Fischer:2008uz,Ellwanger:1995qf,Ellwanger:1996wy,Bergerhoff:1997cv,%
  Pawlowski:2003hq,Fischer:2004uk}. In the following
we use the numerical solutions of the FRG equations for the propagators in
\cite{Fischer:2008uz}, the gluon propagator is displayed in
Fig.~\ref{fig:propplots}.  
%
%
The vertices used for this solution are fully RG-dressed and depend on
one (symmetric) momentum scale $k$. The four gluon vertex is partially
2PI-resummed and hence includes also the sunset diagram with full
propagators. If compared with similar DSE computations, see
\cite{Fischer:2008uz,Alkofer:2000wg,Fischer:2006ub,Binosi:2009qm,Alkofer:2008bs}
and references therein, it is in particular the latter property which
includes contributions that are contained in the two-loop diagrams in
the DSEs. These diagrams are neglected in most approximations to DSEs
(see however \cite{Bloch:2003yu}). This fact is most probably
responsible for the minor deviations of the DSE propagators from the
lattice results,
e.g. \cite{Sternbeck:2006cg,Oliveira:2007dy,Cucchieri:2007md,Bogolubsky:2007ud},
in the mid momentum regime around the peak of the gluon dressing
function in Fig.~\ref{fig:propplots}, for more details see
\cite{Fischer:2008uz}. In spite of this both functional approaches
provide results which are in quantitative agreement with the lattice
propagators.

\subsubsection{Quark confinement from Yang-Mills propagators}

The Polyakov loop variable $L(\vec{x})$, 
\begin{equation}\label{eq:Polloop}
L(\vec x)=\frac{1}{\Nc} \tr\, \text{P}\, \exp { \I g\int_0^\beta d t\, {A}_0(t,\vec x)}\,,
\end{equation}
in QCD with $N_c$ colors and infinitely heavy quarks is related to the
operator that generates a static quark. The trace in 
\eq{eq:Polloop} is evaluated in the fundamental
representation, $\text{P}$ stands for path ordering, and $\beta=1/T$
is the inverse temperature. We can interpret the logarithm of the
expectation value $\langle L\rangle$ as half of the free energy
$F_{q\bar q}$ of a static quark--anti-quark pair at infinite
distance. Moreover, $\langle L\rangle$ is an order parameter for the
center symmetry of the gauge group, see e.g.~\cite{Greensite:2003bk}.

It can be shown that $L[\langle A_0\rangle ]$ also serves as an order
parameter, \cite{Braun:2007bx,Marhauser:2008fz}: we have $L[\langle
A_0\rangle ]\geq \langle L\rangle$ with help of the Jensen
inequality. We also have $L[\langle A_0\rangle ]\equiv 0$ in the
center-symmetric phase. Hence it vanishes exactly at $T_{\rm
  conf}$. Accordingly we have to simply find the solution to the
equation of motion for $A_0$: $V_{\rm YM}'[\langle A_0\rangle
]=0$. The effective potential $V_{\rm YM}[A_0]$ is nothing but the
effective action $\Gamma$ evaluated for constant fields $A_0$, $V_{\rm
  YM}[A_0]=\Gamma[A=A_0,C=0,\bar C=0]$. Its flow is governed by the
ghost and gluon diagrams in Fig.~\ref{fig:funflowQCD}, and hence can
be computed solely from the ($k$-dependent) ghost and gluon
propagators \cite{Braun:2007bx}. In other words, $V_{\rm YM}$ being
confining for low temperatures puts constraints on the behaviour of
the ghost and gluon propagators in the deep infrared. Loosely
speaking, confinement demands a minimal amount of infrared ghost
dominance \cite{Braun:2007bx}. Interestingly, infrared stability in
the background Landau gauge puts an upper bound on the amount of
infrared ghost enhancement \cite{Eichhorn:2010zc}. Together this puts
rather tight constraints on the infrared asymptotics of the
propagators which are satisfied by the actual numerical solutions
\cite{Eichhorn:2010zc}. Note, however, that the critical temperature
is insensitive to the deep infrared \cite{Braun:2007bx,Braun:2010cy},
see also \cite{Fischer:2009wc,Fischer:2010fx}. Moreover, confinement
is not directly sensitive to the size of the coupling. These
statements hold true in dynamical QCD and extends to other observables
\cite{Braun:2009gm}.

\begin{figure}[t]
 \hspace{-.6cm} \includegraphics[height=.235\textheight]{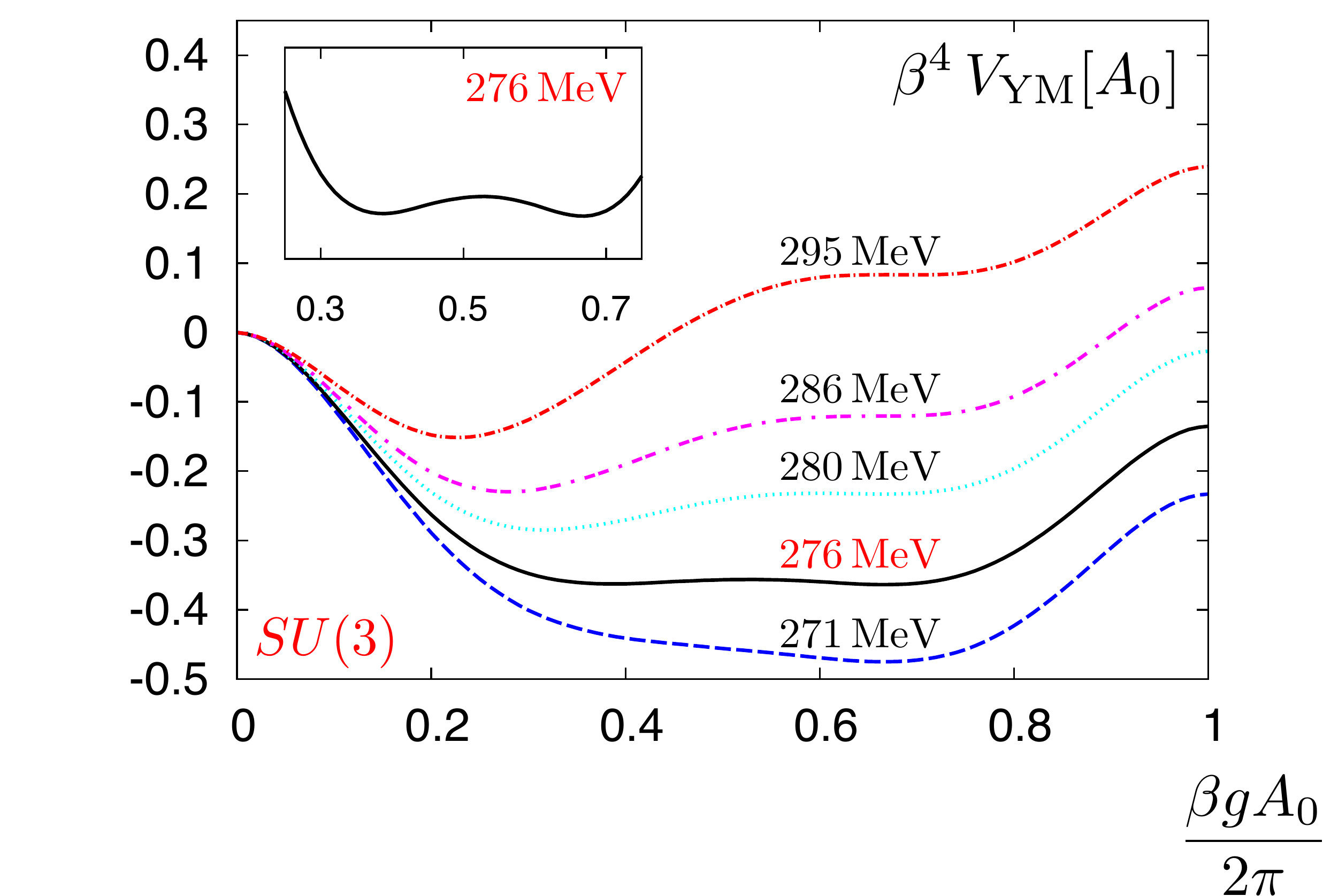}\label{fig:potA}
\caption{Polyakov loop effective potential \cite{Braun:2007bx}. 
 \ \vspace{-.3cm}\ }
\end{figure}

\begin{figure}[b]
  \includegraphics[height=.22\textheight]{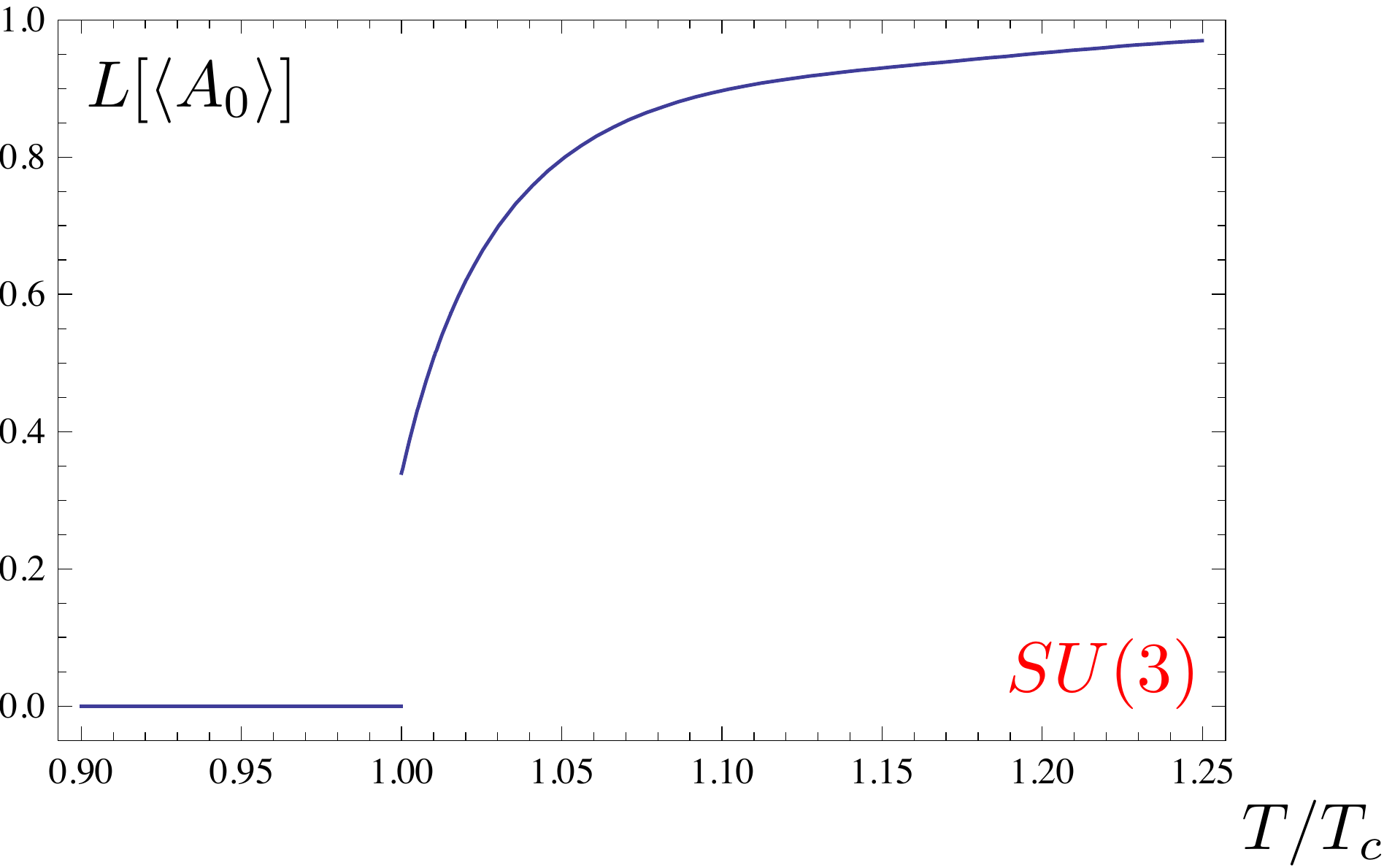}\label{fig:Polloop}
  \caption{Polyakov loop \cite{Braun:2007bx}.\ \vspace{-.4cm}\ } 
\end{figure}

Inserting the propagators shown in Fig.~\ref{fig:propplots} we get the
effective potential shown in Fig.~\ref{fig:potA}.
%
%
The transition temperature is computed as $T_c=276\pm 10$ MeV in
quantitative agreement with the lattice result of $T_c=270$ MeV,
e.g. \cite{Lucini:2003zr}. The order parameter, $L[\langle A_0\rangle
]$ is obtained from the minimum of the potential in
Fig.~\ref{fig:potA} and is shown in Fig.~\ref{fig:Polloop}. If
compared with lattice results for $\langle L\rangle$,
e.g. \cite{Lucini:2003zr}, the Polyakov loop variable $L[\langle
A_0\rangle ]$ shows a steeper slope and hence a smaller transition
region, see also \cite{Dumitru:2010mj}, in accordance with the Jensen
inequality discussed above.  Note, however, that the results presented
in Figs.~\ref{fig:potA} and \ref{fig:Polloop} are obtained by
neglecting the back-reaction of the $A_0$-fluctuations related to
$V_{\rm YM}''[\langle A_0\rangle ]$. These fluctuations reduce the
slope of the Polyakov loop variable $L[\langle A_0\rangle ]$. For a
system at a second order phase transition they carry the universal
properties of a system and drive the system into the symmetric
phase. In turn, they are sub-leading for the critical temperature of a first
order phase transition. This is one of the reasons for the good
quantitative precision of the results for the critical temperature for
$SU(3)$ and higher $SU(N)$ \cite{Braun:2010cy}.

For $SU(2)$ Yang-Mills theory we have a second order phase transition (Ising
universality class), and indeed we find $T_{\rm conf}/\sqrt{\sigma}
=.605$ instead of $T_{\rm conf}/\sqrt{\sigma} =.709$ \cite{Lucini:2003zr}, with string
tension $\sigma$. An alternative computation of the order parameters
in $SU(2)$ has been done in Polyakov gauge. There the flow is
completely described by the $V_{\rm YM}''[A_0]$-fluctuations
\cite{Marhauser:2008fz}, and the critical temperature is computed as $T_{\rm
  conf}/\sqrt{\sigma} =.69$. We also find the Ising class critical
exponents. More recently, the Landau gauge computation has been
extended to the $V_{\rm YM}''[A_0]$-fluctuations with $T_{\rm
  conf}/\sqrt{\sigma} =.705$ \cite{ConfFlucs}. A comparison of the
temperature-dependence of the order parameter in the Polyakov gauge
and the Landau gauge gives a remarkable agreement for all temperatures
and provides strong support for the gauge independence of our results
\cite{Marhauser:2008fz}. This concludes the investigation of
Yang-Mills theory.
%
%

\subsubsection{QCD,  chiral symmetry breaking and dynamical hadronisation}
The matter fluctuations due to dynamical quarks are encoded in the
last two loops in Fig.~\ref{fig:funflowQCD}. The quark-gluon
interaction gives rise to four-fermi terms upon integration of a
momentum shell with the flow. They originate from box diagrams
$\propto$ $\alpha_s^2$, leading to 
\begin{equation}\label{eq:4fermi} 
\int_x \lambda_{\psi,k}(\alpha_s)\left[ (\bar\psi \psi)^2+
(i \bar\psi\gamma_5 \vec\tau\psi)^2\right]
\,,
\end{equation}
where $\tau=(\sigma_1,\sigma_2,\sigma_3)$. Here we have restricted
ourselves to the two-flavour case, $N_f=2$. Further momentum shell
integration then also give contributions to the four-fermi terms
related to \eq{eq:4fermi} as the flow $\partial_t
\lambda_{\psi,k}= \partial_t
\lambda_{\psi,k}(\lambda_\psi,\alpha_s)$. Note that for $\alpha_s=0$
this resembles a NJL-type flow. The four-fermi term in \eq{eq:4fermi}
can be conveniently rewritten with a Hubbard-Stratonovich
transformation as
\begin{equation}\label{eq:Yukawa} 
\012 
\int_x  m_{\sigma,k}^2(\sigma^2+\vec\pi^2) + \int_x  h_k
\left[ (\bar\psi \psi)\,\sigma +
    (i \bar\psi\gamma_5\vec\tau \psi)\,\vec \pi  \right] \,,
\end{equation}
with $m_{\sigma,k}^2=h_k^2/(2 \lambda_{\psi,k})$ and using the
equations of motion for $\sigma$ and $\vec\pi$. Further momentum shell
integrations generate kinetic terms for the mesonic degrees of freedom
$\sigma$ and $\vec\pi$ and further interaction terms, in particular an
effective potential $V_{\rm eff,k}[\sigma^2+\vec \pi^2]$ which
includes the mesonic mass in \eq{eq:Yukawa}. Additionally the
four-fermi interaction \eq{eq:4fermi} is re-generated from the
quark-gluon interaction. This already leads to an effective action
$\Gamma_k[\phi]$ with $\phi=(A,C,\bar C,\psi,\bar\psi;\sigma,\vec
\pi)$, and suffices to describe the onset of chiral symmetry breaking,
\cite{Braun:2009gm,Gies:2002hq,Gies:2005as,Braun:2006jd,Braun:2008pi}. The
above structure has been also advocated in \cite{Kondo:2010ts} on the
basis of the pure glue results in \cite{Marhauser:2008fz}. An
important feature of the above bosonisation is that it is not subject
to double-counting problems. With cut-off terms for the mesonic
degrees we are led to the flow equation \eq{eq:FRG} depicted in
Fig.~\ref{fig:funflowQCD}.

In this setting chiral symmetry breaking is monitored by the
expectation value of $\sigma$: the Yukawa term in \eq{eq:Yukawa} can
be absorbed into the Dirac term in \eq{eq:SQCD} by
\begin{equation}\label{eq:maas+mesons}
 i  \,m_{\psi,k}\bar\psi \psi\to \bar\psi\left(i\, m_{\psi,k}+h_k( \sigma
    + i\gamma_5\vec\tau \pi)\right)\psi\,. 
\end{equation}
Hence, for a non-vanishing expectation value of $\sigma$ this term serves
as an additional mass term for the quarks. We concentrate on the
Yukawa term and the effective mesonic potential at vanishing pion
field, $\vec\pi=0$,
\begin{equation}\label{eq:chiral1}
\int_x V_{\rm eff,k}[\sigma^2]  +\int_x h_k(\bar\psi \psi)\,\sigma\,. 
\end{equation} 
In the symmetric phase the effective potential $V_{\rm eff,k}$ has its
minimum at $\sigma=0$ and the fermionic term vanishes on the equation
of motion. In the broken phase the minimum is at $\bar\sigma\neq 0$,
$V_{\rm eff}'[\bar\sigma^2]=0$ and the fermionic term gives rise to a
mass term with mass $|h_k|\bar\sigma$. Indeed, $f_\pi\simeq
|h_k|\bar\sigma$ is an order parameter for the chiral phase transition
which happens at $m^2_{\sigma}=0$, that is $\lambda_\psi\to
\infty$. Note that even within this simplified setting this already
introduces momentum-dependent (non-local) four-fermi couplings. Via
the coupling to the effective mesonic potential and the kinetic terms 
one also generates (non-local) higher order fermionic terms. 

The above approach is systematically improved by dynamical
hadronisation or rebosonisation
\cite{Pawlowski:2005xe,Gies:2001nw,Gies:2002hq,Floerchinger:2009uf}:
the re-generated four-fermi interaction can be re-absorbed in the
Yukawa-interaction in \eq{eq:Yukawa}. This dynamically re-adjusts the
expansion of the effective action in the scale-dependent relevant
degrees of freedom and guarantees or at least improves the locality of
the expansion in relevant degrees of freedom. In this way the system
evolves dynamically from the high temperature/large cut-off scale
quark-gluon phase to the low temperature/small cut-off scale hadronic
phase, see in particular \cite{Braun:2008pi}.

With or without dynamical hadronisation it is a particular strength of
the present approach that it allows a direct access to the physics
mechanisms. The r$\hat{\rm o}$le of the gauge coupling for chiral
symmetry breaking is already easily displayed in the setting without
(full) dynamical hadronisation. In the perturbative regime of QCD the
four-fermi coupling can be safely put to zero. The aforementioned box
diagrams then generate and successively increase the strength of the
four-fermi coupling. If it exceeds a critical strength, spontaneous
chiral symmetry breaking is induced similarly to the NJL model. In the
present QCD approach, this is triggered by the increase of the gauge
coupling $\alpha_s$: for $\alpha_s<\alpha_{s,\rm crit}$ spontaneous
chiral symmetry breaking is triggered
\cite{Braun:2009gm,Gies:2002hq,Gies:2005as,Braun:2006jd}. In turn, if
$\alpha_s$ does not exceed the critical coupling
$\alpha_s>\alpha_{s,\rm crit}$, the four-fermi coupling never grows
big and runs into the Gau\ss ian infrared fixed point $\lambda_\psi=0$
with chiral symmetry. We conclude that chiral symmetry breaking is
primarily driven by the strength of the gauge coupling in
contradistinction to the deconfinement transition.

\subsubsection{Results: phase structure at vanishing density}

The approach described in the preceding sections is now put to work in
two flavour QCD in the chiral limit and at vanishing density. We map
out the confinement-deconfinement phase transition with the Polyakov
loop order parameter $L[\langle A_0\rangle ]$, and the pion decay
constant $f_\pi\simeq \langle \sigma\rangle$. The
confinement-deconfinement phase transition is also accessible by
so-called dual order parameters which are derived from quark
correlation functions
\cite{Fischer:2009wc,Gattringer:2006ci,Synatschke:2007bz,Bilgici:2008qy,Zhang:2010ui}
with non-trivial temporal boundary conditions for the quark fields in
the chosen correlator. In \cite{Braun:2009gm} it has been shown that
such order parameters can be also derived from QCD${}_\theta$ at
imaginary chemical potential $\mu=2 \pi i\,\theta/\beta$, for more
details see also \cite{Haas:2010bw} in these proceedings. Here we only
remark that imaginary chemical potential can be recast as a
non-trivial boundary condition for the quarks,
\begin{equation}\label{eq:imagchem}
\int \bar\psi_\theta\left( i \dr
  +i\,m_\psi\right)\psi_\theta\,,\quad \psi_\theta(t+\beta,\vec x)=- e^{2
  \pi i\theta} \psi(t,\vec x)\,.
\end{equation} 
It can be shown that the first Fourier moment in $\theta$ of {\it any} observable
is sensitive to center symmetry \cite{Braun:2009gm,Haas:2010bw}.
Hence, if it does not vanish identically, it is an order parameter for
the confinement-deconfinement phase transition. 
This allows to define
the so-called dual pressure (or density) as an order parameter for the
confinement-deconfinement phase transition (in pure Yang-Mills),
\begin{equation}\label{eq:dualpressure} 
  \tilde p=\int_0^\theta d\theta\, e^{-2 \pi i\theta}\, 
\Gamma_\theta[\langle A_0\rangle_{\theta=0},
\langle \sigma\rangle_{\theta=0}]\,.  
\end{equation}
Here, $\Gamma_\theta$ is the effective action of QCD${}_\theta$, and
$\langle A_0\rangle_{\theta=0}$ and $\langle
\sigma\rangle_{\theta=0}$ are the solutions of the equations of motion in
the physical theory at $\theta=0$. The advantage of the order
parameter \eq{eq:dualpressure} is, that its flow $\partial_t \tilde
p$ is directly related to the flow of the effective action, 
\eq{eq:FRG}. Clearly, this is the object least sensitive to a given
approximation of $\Gamma_k$.
%
\begin{figure}[t]
  \includegraphics[height=.27\textheight]{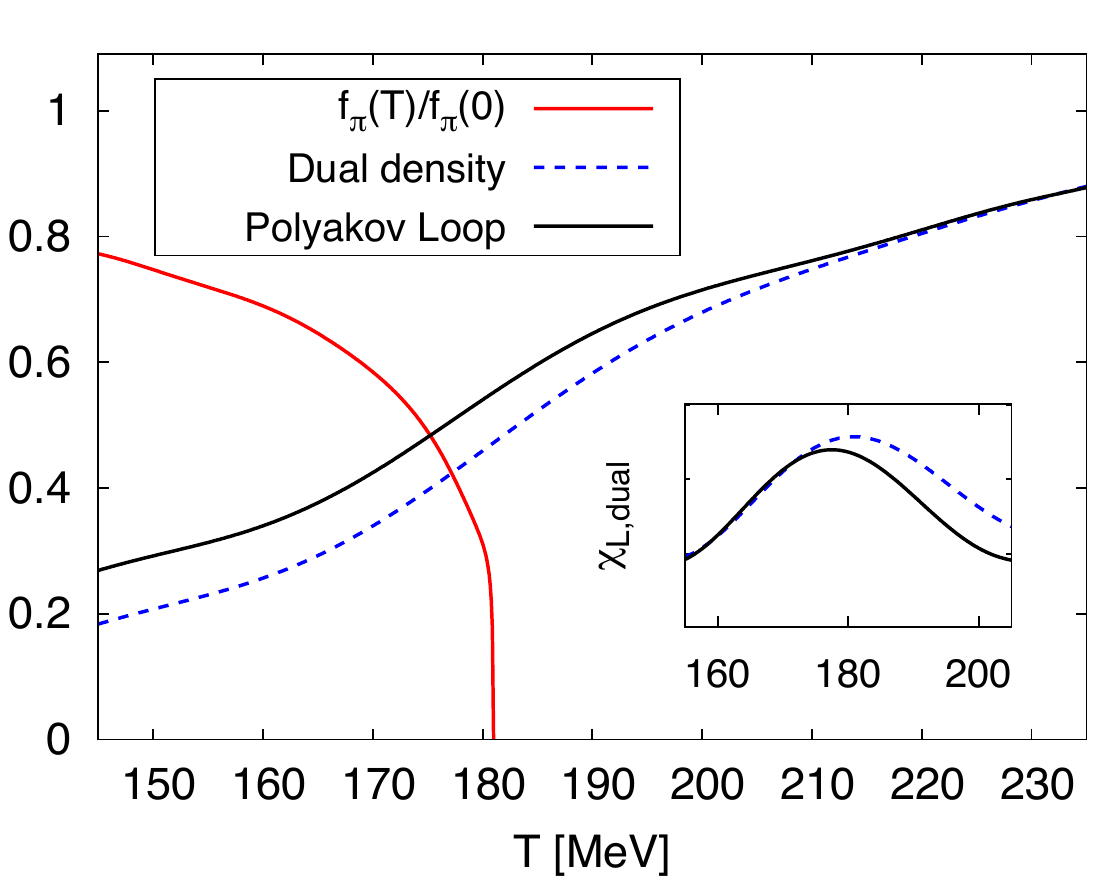}\label{fig:phase-structure}
  \caption{Pion decay constant, dual pressure difference and
    Polyakov loop as functions of temperature, $\chi_L=\partial_T
    L$, $\chi_{\text{dual}}=\partial_T \tilde{p}$ \cite{Braun:2009gm}.\ \vspace{-.0cm}\ }
\end{figure}
%

The results for the order parameters are displayed in
Fig.~\ref{fig:phase-structure} and suggest a relation between the
chiral phase transition and the broad confinement-deconfinement
cross-over with a width of approximately 20 MeV, both happen at about
$180$ MeV. The agreement between the cross-over temperature derived
from the Polyakov loop and the dual pressure is remarkable as is the
apparent close relation of their values for all temperatures. The
latter has been studied in more detail in
\cite{Braun:2009gm,Haas:2010bw} and we defer the reader to this
work. Beside its formal relation it constitutes a non-trivial
consistency check of the approximation used as the Polyakov loop is
dominated by gluonic fluctuations whereas the dual pressure is
dominated by matter fluctuations. Finally we would like to remark that
qualitatively our findings compare well with the lattice findings for
2+1 flavours, see
e.g. \cite{Aoki:2009sc,Cheng:2009zi,Borsanyi:2010zi,Bazavov:2010pg,Kanaya:2010qd}.

\subsubsection{Results: phase structure at finite density}
A first step towards non-vanishing density is the inclusion of an
imaginary chemical potential. It allows us to compute the dual order
parameters such as the dual pressure \eq{eq:dualpressure}. We also can
collect indirect information about real chemical potential by
continuation. In specific cases this may allow to fix the phase
structure at real chemical potential \cite{deForcrand:2010he}.

\begin{figure}[t]
  \includegraphics[height=.3\textheight]{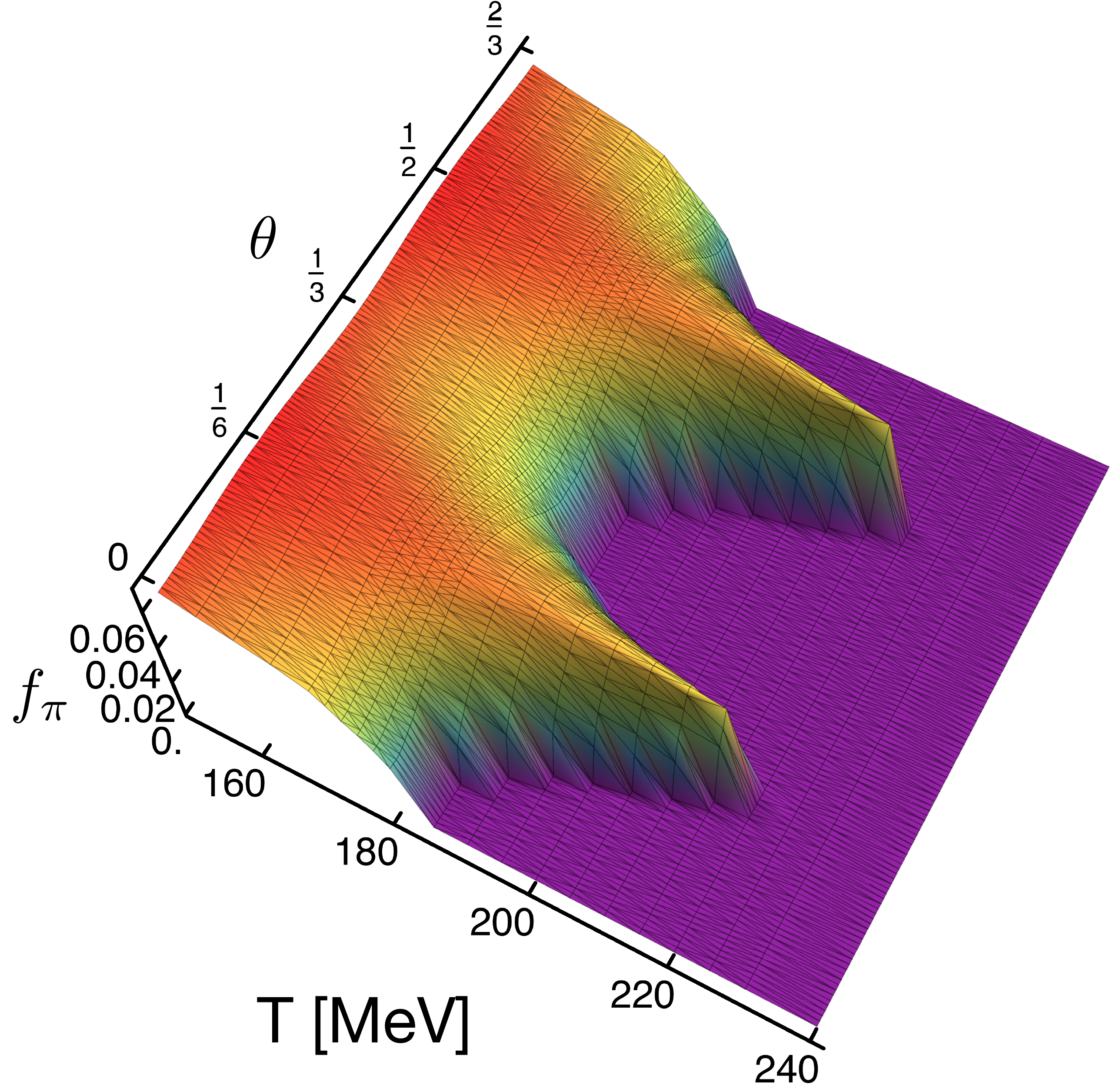}\label{fig:fPi3d}
  \caption{Pion decay constant as a function of imaginary chemical
  potential and temperature \cite{Braun:2009gm}.\ \vspace{-.5cm}\ }
\end{figure}

\begin{figure}[t]
  \includegraphics[height=.23\textheight]{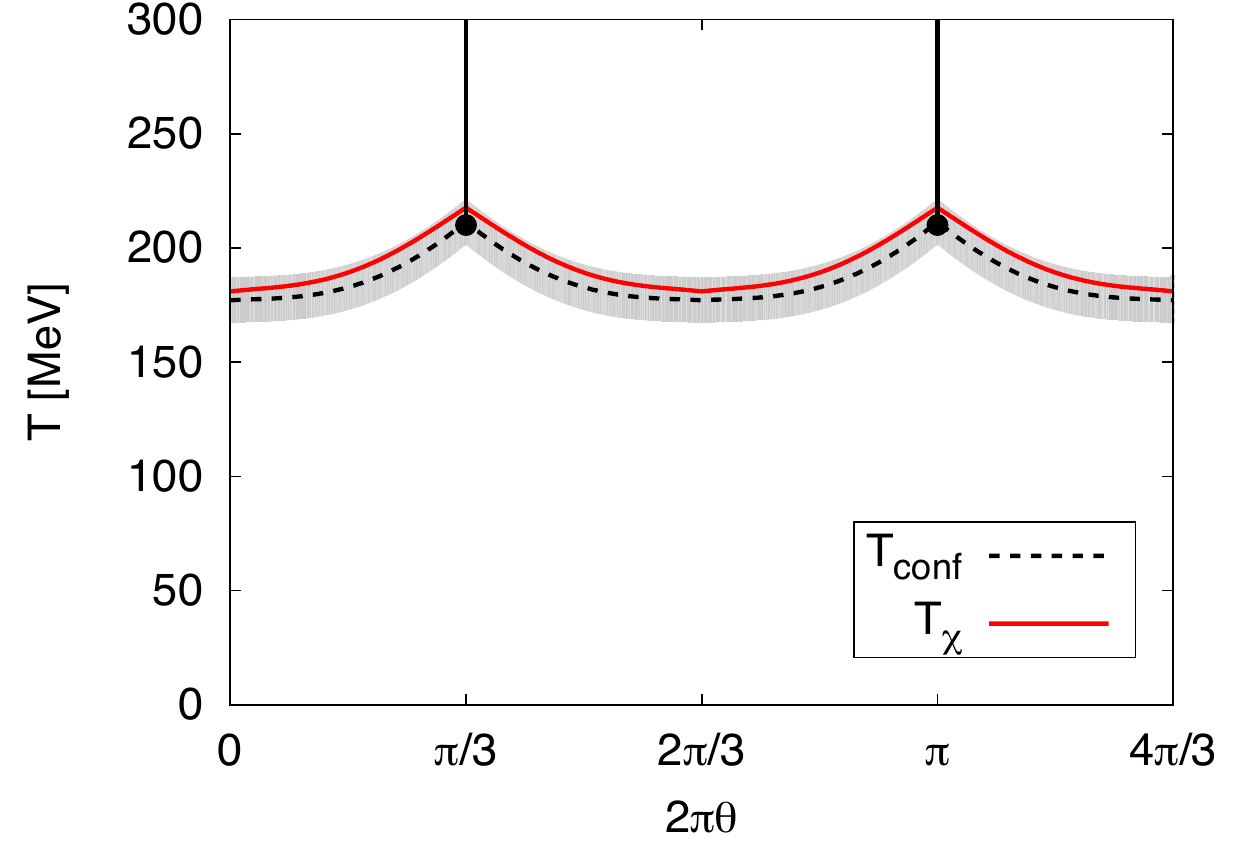}\label{fig:phasediagram-immu}
  \caption{QCD phase diagram at imaginary chemical potential. The
    grey band represents the width of $\chi_L$. Black dots indicate the endpoints of the 
  Polyakov loop RW transitions \cite{Braun:2009gm}.\ \vspace{-0cm}\ }
\end{figure}

The pion decay constant is displayed in Fig.~\ref{fig:fPi3d}. 
%
%
Together with the Polyakov loop it leads to the phase structure
displayed in Fig.~\ref{fig:phasediagram-immu}.  We conclude that the
close relation between the chiral phase transition and the
confinement-deconfinement cross-over persists at imaginary chemical
potential. This is in agreement with related lattice computations at
relatively heavy quark masses,
e.g. \cite{deForcrand:2002ci,D'Elia:2002gd}. In contradistinction, in
model computations this is only seen within an adjustment of the
coupling of eight-fermi interaction \cite{Sakai:2008py}. In the
present QCD approach these terms with their QCD-induced coupling are
generated by the flow, no adjustment is required. This pattern allows
us to access the interesting question of the nature of the
Roberge-Weiss endpoint. This is important for the extension of the
computations at imaginary chemical potential to real chemical
potential, \cite{deForcrand:2010he,D'Elia:2009qz}, and is currently
under investigation in the present approach.
%
%

Finally we consider real chemical potential. As has been stressed
before, functional approaches do not suffer from the sign problem that
is so virulent within lattice simulations at real chemical
potential. A FRG study of the chiral phase boundary in one-flavour QCD
has been put forward in \cite{Braun:2008pi}. The results include in
particular the curvature of the chiral phase boundary which is
consistent with lattice results for small chemical potential.

First computations within two-flavour QCD at real chemical potential
have also been performed. Most notably the two critical temperatures
stay close to each other; indeed, the confinement-deconfinement
temperature tends to get smaller in comparison to the chiral one. This
is not seen in most computations with Polyakov loop-extended
models and is related to the missing back-coupling of the matter
sector into the glue sector in these models. To see this more clearly
we use the fact that the above-mentioned models,
e.g. \cite{Schaefer:2007pw,Fukushima:2003fw,Ratti:2005jh,Sakai:2008py}
and their non-local versions, e.g. \cite{Hell:2008cc}, can be
interpreted as specific approximations of the fully dynamical QCD flow
in \cite{Braun:2009gm}, for more details on this relation see also \cite{Herbst:2010rf}. We
first remark that the matter sector including fluctuations is given by
the last two matter diagrams in Fig.~\ref{fig:funflowQCD}. Switching
off the glue contributions to the matter propagators leads to general
quark-meson models which have been studied intensively beyond mean
field with the FRG, for reviews see
e.g. \cite{Berges:2000ew,Schaefer:2004en,Schaefer:2006sr}.

In turn, the glue part of QCD is encoded in the first two, pure glue,
diagrams. Switching off the matter contributions to the glue dynamics
reduces the glue part to Yang-Mills theory, and the Polyakov loop
potential reduces to that of Yang-Mills theory. In the Polyakov
loop-extended models, the coupling of these two, now decoupled,
sectors is reintroduced via the $A_0$- (or Polyakov loop $L$) and
$\sigma$-background dependence of the quark loop. Note that such an
approximation of the QCD flow in Fig.~\ref{fig:funflowQCD} still
involves a fluctuating matter sector. This describes these models
beyond mean field as has been studied in
\cite{Herbst:2010rf,Skokov:2010wb,Skokov:2010uh}.
%
\begin{figure}[t]\ \hspace{-.3cm} 
  \includegraphics[height=.255\textheight]{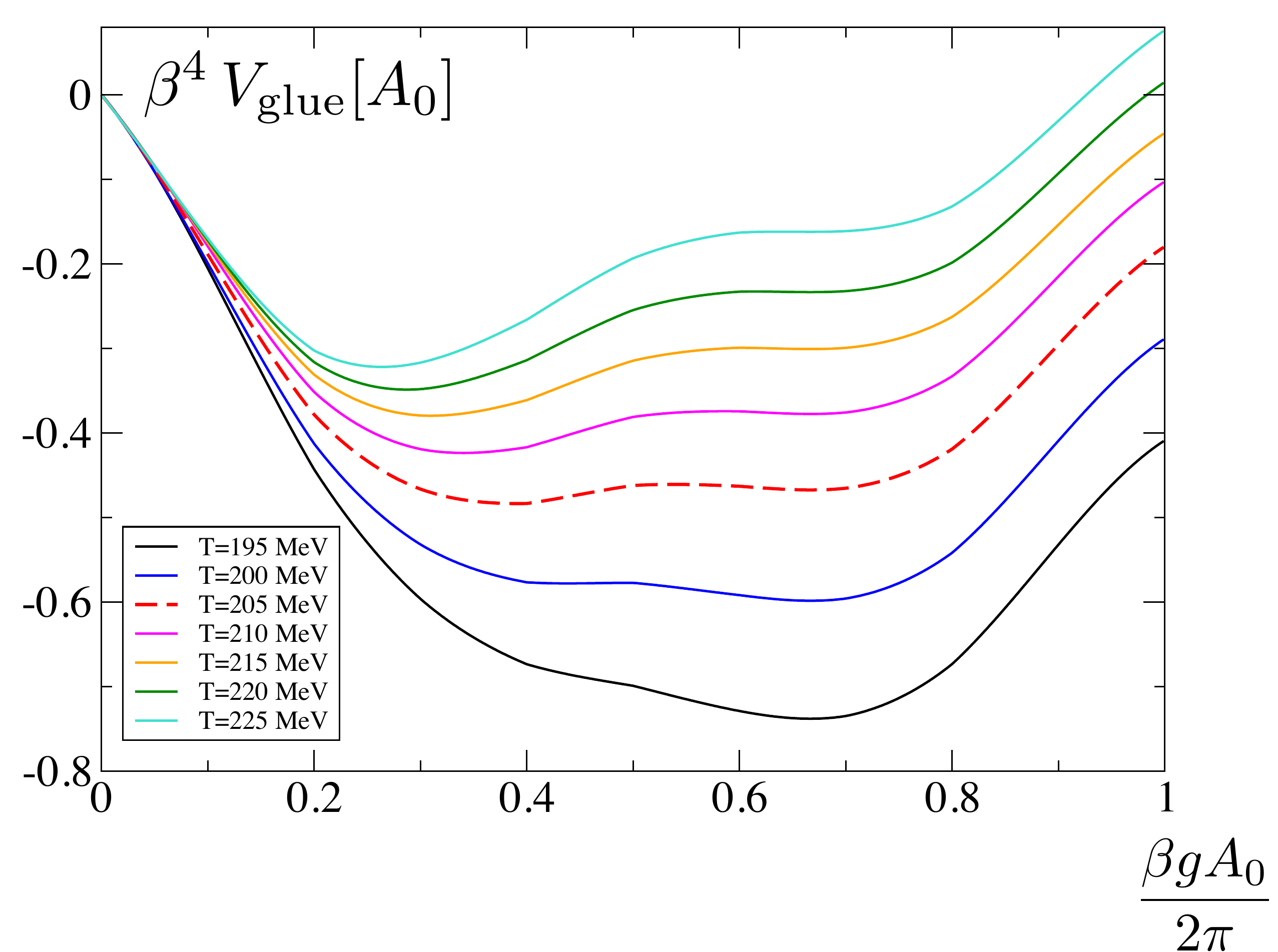}\label{fig:pureglue}
  \caption{Pure glue part of the Polyakov loop potential in two-flavour QCD.\ \vspace{-0cm}\ }
\end{figure}
%

The above connection of the Polyakov loop-extended models to the
present QCD approach can be used to qualitatively improve these
models: the full QCD effective potential (or grand potential) $V_{\rm
  QCD}$ is obtained by evaluating the integrated flow in
Fig.~\ref{fig:funflowQCD} for constant $A_0$- and $\sigma$-fields \cite{Braun:2009gm}, 
\begin{equation} 
  V_{\rm QCD}=V_{\rm glue}[A_0,\sigma]
+V_{\rm quark}[A_0,\sigma]+V_{\rm meson}[A_0,\sigma]\,. 
\end{equation} 
The glue potential $V_{\rm glue}$ is computed from the ghost and gluon
loop in Fig.~\ref{fig:funflowQCD}, $V_{\rm quark}$ is computed from
the quark loop, and $V_{\rm meson}$ is computed from the mesonic
loop.  $V_{\rm quark}+V_{\rm meson}$  are the
direct matter contributions to the full Polyakov loop potential. 
We emphasise that the propagators in this computation are that
of fully-coupled QCD with dynamical quarks and mesons
\cite{Braun:2009gm}. 

We have already remarked that $V_{\rm glue}$ is approximated by the
Yang-Mills potential $V_{\rm YM}$ in the Polyakov loop-extended
models. With $V_{\rm glue}$ from the QCD flows in \cite{Braun:2009gm}
this approximation can be resolved which will be detailed
elsewhere. Here we discuss how to properly adjust the parameters in
the Yang-Mills potential shown in Fig.~\ref{fig:potA} by comparing it
to the full glue potential. For this purpose we evaluate $V_{\rm
  glue}$ on the solution $\sigma_0$ of the equations of motion,
$V_{\rm glue}[A_0]=V_{\rm glue}[A_0,\sigma_0]$ as shown in
Fig.~\ref{fig:pureglue}. It turns out that the related Polyakov loop
compares well with the Yang-Mills one in Fig.~\ref{fig:Polloop} if
both are plotted as functions of $T/T_c$.  Indeed, the form of the
pure glue potential is qualitatively unchanged in comparison to the
Yang-Mills potential, and still shows a first order phase
transition. The latter has to be taken with a grain of salt as we have
dropped the center-breaking, though sub-leading, $V_{\rm QCD}''$-terms
in the gluon propagator. In any case, the transition temperature is
significantly reduced, from $T_{\rm YM}=276$ MeV to $T_{\rm
  glue}\approx 205$ MeV.  We conclude that the full glue potential is
modeled well by the Yang-Mills potential with a reduced critical
temperature. We also remark that the above procedure allows in general
to adjust the sensitive parameters of the Polyakov loop-extended
models with QCD-input. Most importantly, this allows for a systematic
improvement of these models towards full QCD. 

\begin{figure}
\includegraphics[height=.245\textheight]{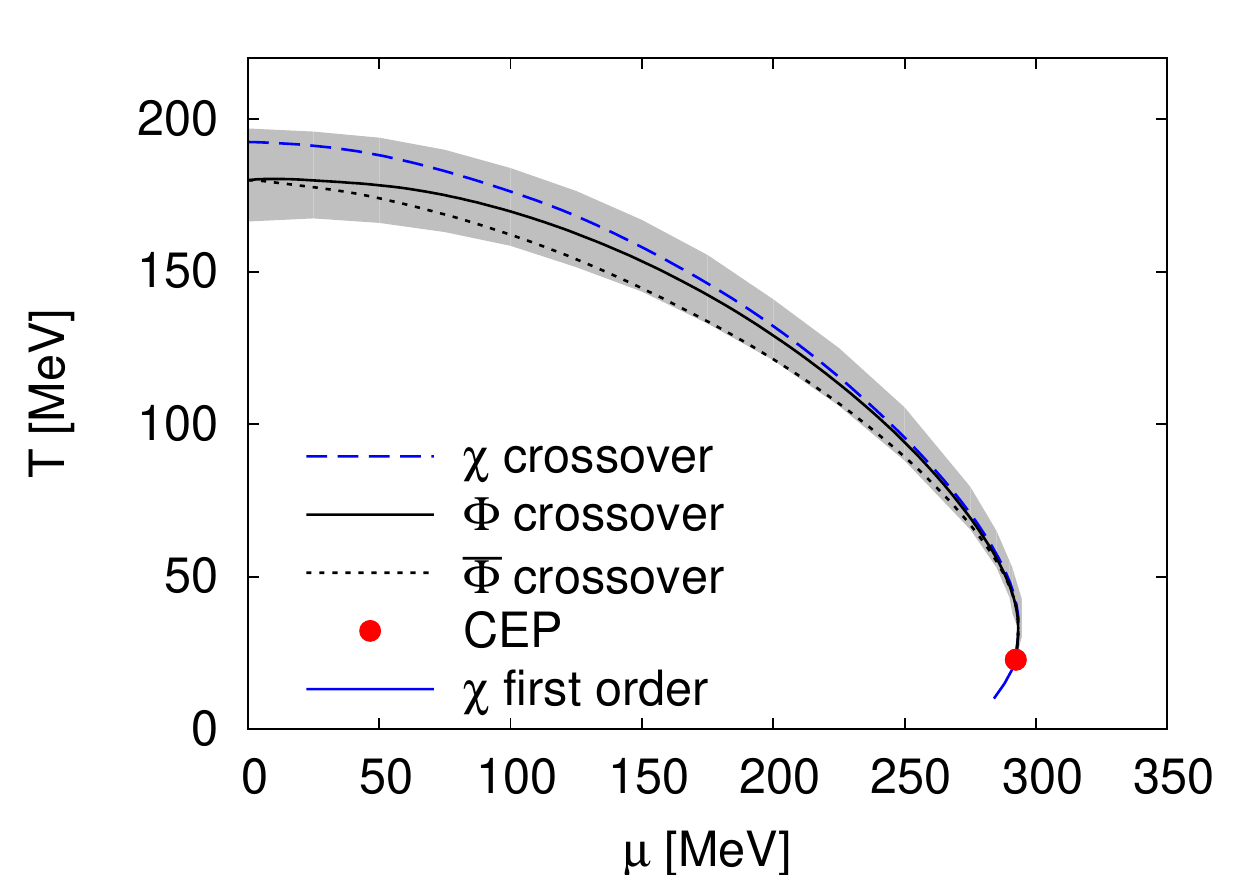}\label{fig:PQM}
\caption{\label{fig:phasediagram-realmu} Chiral and deconfinement
  phase diagram in the PQM model. The (grey) band corresponds to the
  width of $d\Phi/dT$ at $80\%$ of its peak height
  \cite{Herbst:2010rf}.\ \vspace{-.6cm}\ }
\end{figure}

The above analysis confirms quantitatively the
phenomenological HTL/HDL estimate in \cite{Schaefer:2007pw}. Using the
latter also for larger chemical potential as well as utilising FRG
flows for the matter fluctuations in the presence of a Polyakov loop
we are led to the phase diagram displayed in Fig.~\ref{fig:PQM}, see
\cite{Herbst:2010rf}.
At large densities or large chemical potential the present
approximation without baryons certainly is not trustworthy, in
particular in the hadronic phase. The important inclusion of
baryons shall be put forward with dynamical hadronisation. Nonetheless
the combined results from both the model computations and the QCD
flows constrains the position of the critical point; a conservative
estimate leads to $\mu/T> 2/3$.

\subsubsection{Conclusions}
In summary we have put forward a quantitative functional approach to
the phase diagram of QCD at finite temperature and density. So far
fresh insight has been gained in Yang-Mills theory, in one-flavour and
in two-favour QCD at finite temperature and density. The results
compare well to lattice computations without the fixing of additional
parameters. However, the approach also provides results beyond the
applicability range of lattice simulations, in particular at large
density. Currently it is extended to 2+1 flavours as well as to baryons.



\paragraph{Acknowledgments}
I thank the organisers for hosting such a stimulating conference.  I
thank J.~Braun, A.~Eichhorn, C.~S.~Fischer, L.~Fister, H.~Gies, 
L.~M.~Haas, T.~K.~Herbst, A.~Maas, F.~Marhauser, B.-J.~Schaefer, F.~Spallek and
L.~von Smekal for discussions and a pleasant collaboration on the work
presented here. This work is supported by Helmholtz Alliance HA216/
EMMI.



\bibliographystyle{aipproc}   

\bibliography{bibliography}
\vfill
\IfFileExists{\jobname.bbl}{}
 {\typeout{}
  \typeout{******************************************}
  \typeout{** Please run "bibtex \jobname" to optain}
  \typeout{** the bibliography and then re-run LaTeX}
  \typeout{** twice to fix the references!}
  \typeout{******************************************}
  \typeout{}
 }
\end{document}